\documentclass[aps,twocolumn]{revtex4}%
\usepackage{amsfonts}
\usepackage{amsmath}
\usepackage{amssymb}
\usepackage{graphicx}%
\begin{document}
\title{Dynamic properties of a diluted pyrochlore cooperative paramagnet (Tb$_{p}%
$Y$_{1-p}$)$_{2}$Ti$_{2}$O$_{7}$}
\author{A.~Keren$^{1,2}$, J. S. Gardner$^{3}$, G. Ehlers$^{4}$, A. Fukaya$^{5}$, E.
Sagi$^{1}$, and Y.~J.~Uemura$^{5}$}
\affiliation{$^{1}$Department of Physics, Technion - Israel Institute of Technology, Haifa
32000, Israel. $^{2}$Rutherford Appleton Laboratory, Chilton Didcot,
Oxfordshire OX11 0QX, U.K. $^{3}$Brookhaven National Laboratory, Upton, New
York 11973-5000 and NIST Center for Neutron Research, National Institute of
Standards and Technology, Gaithersburg, Maryland 20899-8562. $^{4}$Institut
Laue-Langevin, 6 rue J. Horowitz, 38042, BP 156-98042 Grenoble Cedex 9,
France. $^{5}$Physics Department, Columbia University, New York City, New York
10027, U.S.A.}
\pacs{PACS number}

\begin{abstract}
Investigations of the spin dynamics of the geometrically frustrated pyrochlore
(Tb$_{p}$Y$_{1-p}$)$_{2}$Ti$_{2}$O$_{7}$, using muon spin relaxation and
neutron spin echo, as a function of magnetic coverage $p$, have been carried
out. Our major finding is that paramagnetic fluctuations prevail as
$T\rightarrow0$ for all values of $p$, and that they are sensitive to
dilution, indicating a cooperative spin motion. However, the percolation
threshold $p_{c}$ is not a critical point for the fluctuations. We also find
that the low temperatures spectral density has a $1/f$ behavior, and that
dilution slows down the spin fluctuations.

\end{abstract}
\date{\today}
\maketitle

The study of geometrically frustrated magnetic systems has introduced
important new concepts to condensed matter physics including:
order-by-disorder \cite{STBJAP,ChampionPRB03}, spin ice
\cite{AndersonPR56,HarrisPRL97,RamirezNature99,BramwellScience01}, frustration
driven distortion \cite{KerenPRL01}, and more. A recent and important addition
is the concept of cooperative paramagnetism, whereby an interacting spin
system dynamically fluctuates even as $T\rightarrow0$ \cite{GardnerPRL99b}.
Cooperative paramagnetism on the pyrochlore lattice was first introduced in
the study of Tb$_{2}$Ti$_{2}$O$_{7}$, where despite a substantial
antiferromagnetic Curie-Weiss temperature of $20~$K, a large Tb$^{3+}$ moment
of $9.4~\mu_{B}$ fluctuates continuously down to $T=50$~mK. Bulk and local
probes have been used to study the magnetic nature and the lack of a phase
transition in Tb$_{2}$Ti$_{2}$O$_{7}$\cite{GardnerPRL99b,GingrasPRB00}, and
the results have been interpreted in terms of a cooperative spin system.
However a unique experimental identification that the motion of one spin in
Tb$_{2}$Ti$_{2}$O$_{7}$ influences the motion of its neighbors, is still
lacking. Demonstrating conclusively that the fluctuations in Tb$_{2}$Ti$_{2}%
$O$_{7}$ are indeed a result of spin interactions, and testing to what extent
the fluctuations are cooperative, is the main aim of this work.

Using muon spin relaxation ($\mu$SR) and neutron spin echo (NSE) we
investigate the dynamic fluctuations on a wide time range in (Tb$_{p}$%
Y$_{1-p}$)$_{2}$Ti$_{2}$O$_{7}$ as a function of $p$. Here the non-magnetic Y
ion replaces the magnetic Tb ion on the corner sharing tetrahedra which make
up the pyrochlore lattice \cite{Subramanian}. We vary $p$ from $1$, namely, a
pure pyrochlore, to $0.21$, which is below the percolation threshold
$p_{c}=0.39$ \cite{HenelyCJP01}. $\mu$SR and NSE allow us to determine the $p$
dependence of the waveform and parameters for two important correlation
functions (which are related). These are the auto correlation function
$\left\langle \mathbf{S}_{i}(t)\mathbf{S}_{i}(0)\right\rangle $ where
$\mathbf{S}_{i}$ is the spin operator on the $i^{\prime}$th site, and the
intermediate scattering function $\mathbf{S}(\mathbf{q},t)$ where $\mathbf{q}$
is the wavevector. If interactions are important, we expect the dynamic
properties to be sensitive to $p$. Moreover, if an infinite number of spins
participate in the fluctuations we expect a dramatic change of the correlation
functions at $p_{c}$ since at lower concentration the formation of an
infinitely large island of spins is not possible. Our major finding is that,
indeed, the spin fluctuations in Tb$_{2}$Ti$_{2}$O$_{7}$ result from
interactions. However $p_{c}$ is not a critical point, suggesting that spin
groups of finite size fluctuate cooperatively. We also found that the time
dependent part of the correlation function is a power law.%

\begin{figure}
[ptb]
\begin{center}
\includegraphics[
height=3.6625in,
width=3.0147in
]%
{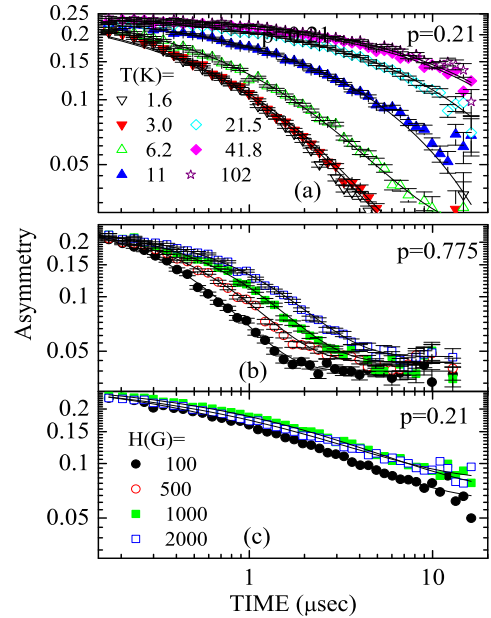}%
\caption{(a) The muon asymmetry versus time at different temperatures for the
most diluted sample. Between 102 and 42~K and below 3~K there is hardly any
temperature dependence. (b) and (c) The muon asymmetry versus time at
$T=100$mK and various different fields for two samples. In (c) the error bars
are omitted for clarity. The fits are to Eq.~\ref{FitFunc} as described in the
text.}%
\label{AsyvsTandH}%
\end{center}
\end{figure}

Our powder samples were made by the standard solid state method described
elsewhere \cite{GardnerPRL99b} and characterized by room temperature X-ray
diffraction and temperature dependent magnetization. First we report on our
$\mu$SR experiments, which were conducted at the ISIS pulsed muon facility,
Rutherford Appleton Laboratory, UK. In these experiments we measure the time
and field dependent asymmetry $A(H,t)$ of the decay positrons from muons
implanted in the sample. This asymmetry is proportional to the muon
polarization $P(H,t)$ along the $\widehat{\mathbf{z}}$ direction, which is the
direction of the external field $H$. In Fig.~\ref{AsyvsTandH} we present the
asymmetry under three different conditions. The log-log plot is used to
emphasize the difference in the asymmetries between different runs. In panel
(a) we depict the muon asymmetry versus time at different temperatures for the
most diluted sample. Note that between $102$ and $42$ K, and below $3$ K, the
muon polarization is almost temperature independent. In panel (b) and (c) we
show the time dependence of the muon asymmetry at $T=100$~mK for two different
samples in various applied fields. At $p=0.775$ the muon relaxation clearly
decreases with increasing field. At a lower concentration $p=0.21$ the
relaxation is lower than that at $p=0.775$ and again decreases between 100 and
1000G, but then saturates. In panel (c) the error bars are omitted for
clarity. Temperature and field dependent raw data for the pure sample (which
agree with ours) can be found in Ref.~\cite{GardnerPRL99b}. In all cases the
raw data are fitted to a stretched exponential relaxation function
\begin{equation}
A(H,t)=A_{0}\exp\left[  -\left(  t/T_{1}\right)  ^{\beta}\right]  +B_{g}
\label{FitFunc}%
\end{equation}
where $B_{g}$ is a background signal representing muons that missed the
sample, $A_{0}$ is the initial asymmetry, $T_{1}$ is the spin lattice
relaxation time, and $\beta$ is the stretching exponent. The fits were done
with a common $\beta$ for all temperature and fields for a given sample. The
solid lines in Fig.~\ref{AsyvsTandH} are from these fits. It is important to
mention that $T_{1}$ obtained by such a fit is identical to that determined by
the $1/e$ criteria ($A(H,T_{1})=A(H,0)/e$).%

\begin{figure}
[ptb]
\begin{center}
\includegraphics[
natheight=8.285800in,
natwidth=10.892300in,
height=2.8772in,
width=3.774in
]%
{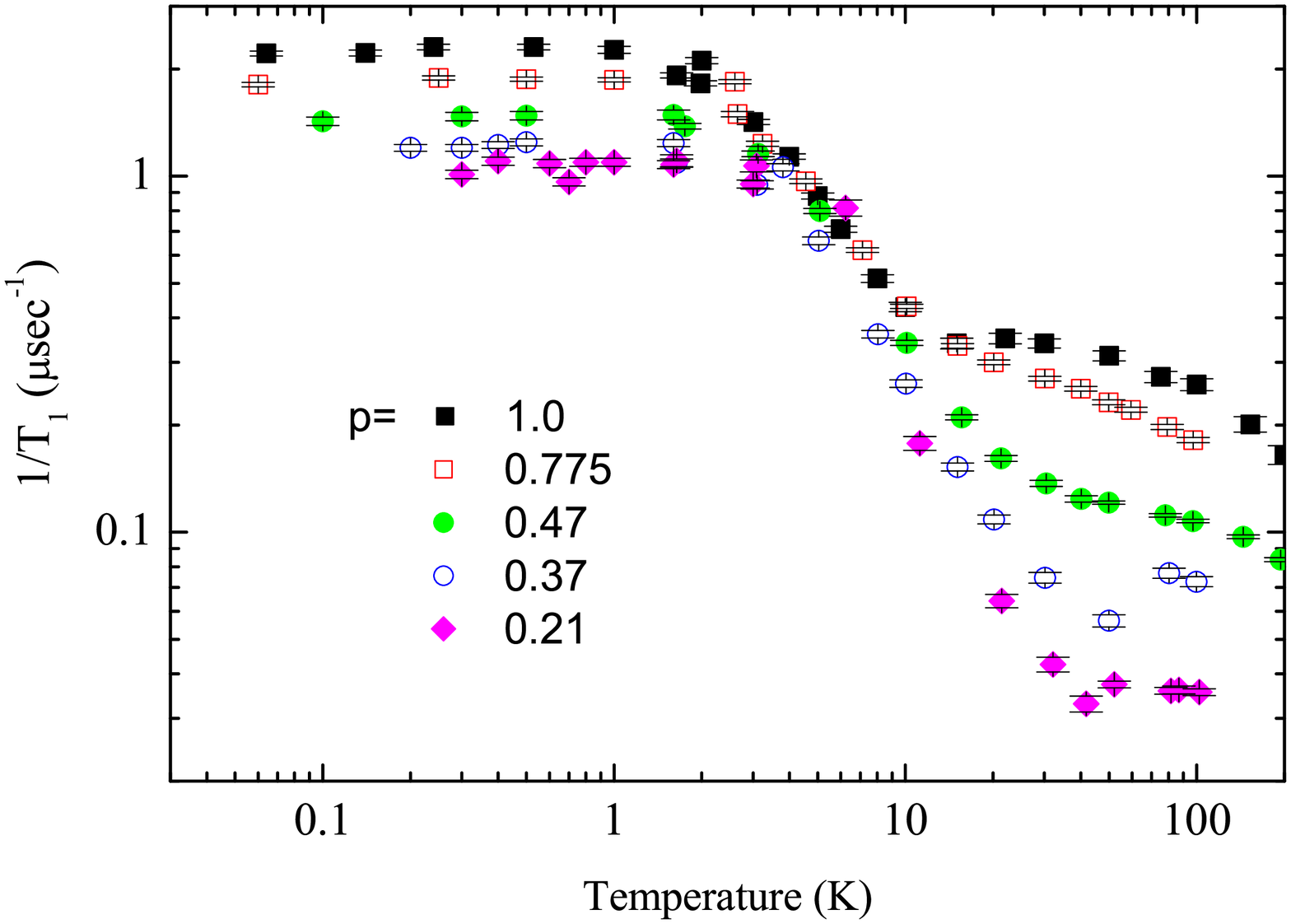}%
\caption{Temperature dependence of the muon relaxation rate $1/T_{1}$ at
H=50G.}%
\label{T1vsT}%
\end{center}
\end{figure}

The muon relaxation rate $1/T_{1}$ obtained from the fits at $H=50$ G as a
function of temperature for several samples is presented in Fig.~\ref{T1vsT}.
The observed temperature dependence of $1/T_{1}$ is very unusual. First of all
$1/T_{1}$ shows no $T$ dependence at the lowest $T$ region for all $p$. This
behavior is usually associated with paramagnetism, and it is observed in
several frustrated magnets \cite{UemuraPRL,KerenPRL00,HodgesPRL02}, low
dimensional systems with singlet ground state and dangling bonds
\cite{KojimaPRL95}, and in high spin molecules \cite{SalmanPRB02}. Most
magnets show a $1/T_{1}$ maximum at some critical temperature. Second, the
slowing down of spin fluctuations occurs in two steps, one at high
temperatures, followed by a plateau around $50$~K, and another step starting
around $\sim20$~K. This might be a result of populations of a crystal field
doublet which is $18$~K above the ground state doublet \cite{GingrasPRB00}.
Finally, both at very high and very low temperatures there are significant
differences in $1/T_{1}$ between samples, however at $4~$K, $1/T_{1}$ of all
samples is the same. This is very intriguing and at present we have no
explanation for this result. However, crystal field levels must be part of
this explanation. In contrast, the $p$ dependence of $1/T_{1}$ below $4$~K is
an indication that correlations are again playing a role in this low
temperature region.

\begin{figure}
[ptb]
\begin{center}
\includegraphics[
natheight=8.078200in,
natwidth=10.864600in,
height=2.8781in,
width=3.8623in
]%
{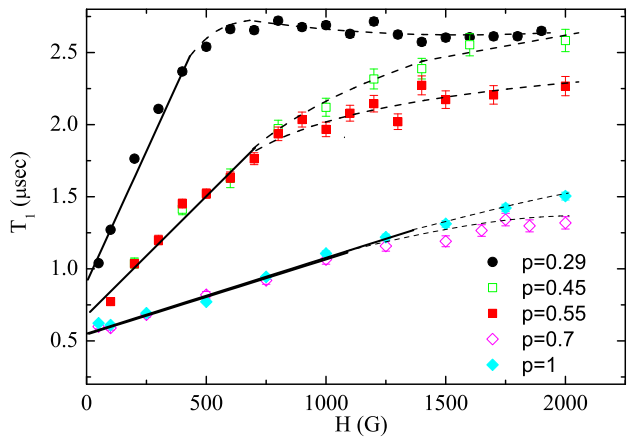}%
\caption{The muon relaxation time $T_{1}$ as a function of the longitudinal
field. The solid lines are linear fits of the low field region at $T=100$~mK.
$dT_{1}/dH$ is obtained from the fits.}%
\label{T1vsH}%
\end{center}
\end{figure}

In Fig.~\ref{T1vsH} we present the muon relaxation time $T_{1}$ at $T=100$~mK
as a function of the longitudinal field for five samples. In all cases $T_{1}$
increases linearly as a function of fields at low fields. The solid lines are
linear fits to the low field region. This region is not affected by diluting
the pure system by 30\%, but further dilution has a strong impact on
$T_{1}(H)$. The dashed lines are continuations of the linear fits and serve as
guides to the eye. The range in which $T_{1}(H)$ is linear decreases with
decreasing $p$, and at low concentrations $T_{1}$ clearly saturates at
$H\sim0.5$~kG. More importantly, the more diluted the sample, the greater is
the sensitivity of $T_{1}$ to $H$. This means that diluting the magnetic
lattice slows down the spin fluctuation since the greatest sensitivity of
$T_{1}$ to external field is when the spin system is static.%

\begin{figure}
[ptb]
\begin{center}
\includegraphics[
natheight=8.073000in,
natwidth=11.584200in,
height=2.8772in,
width=3.4506in
]%
{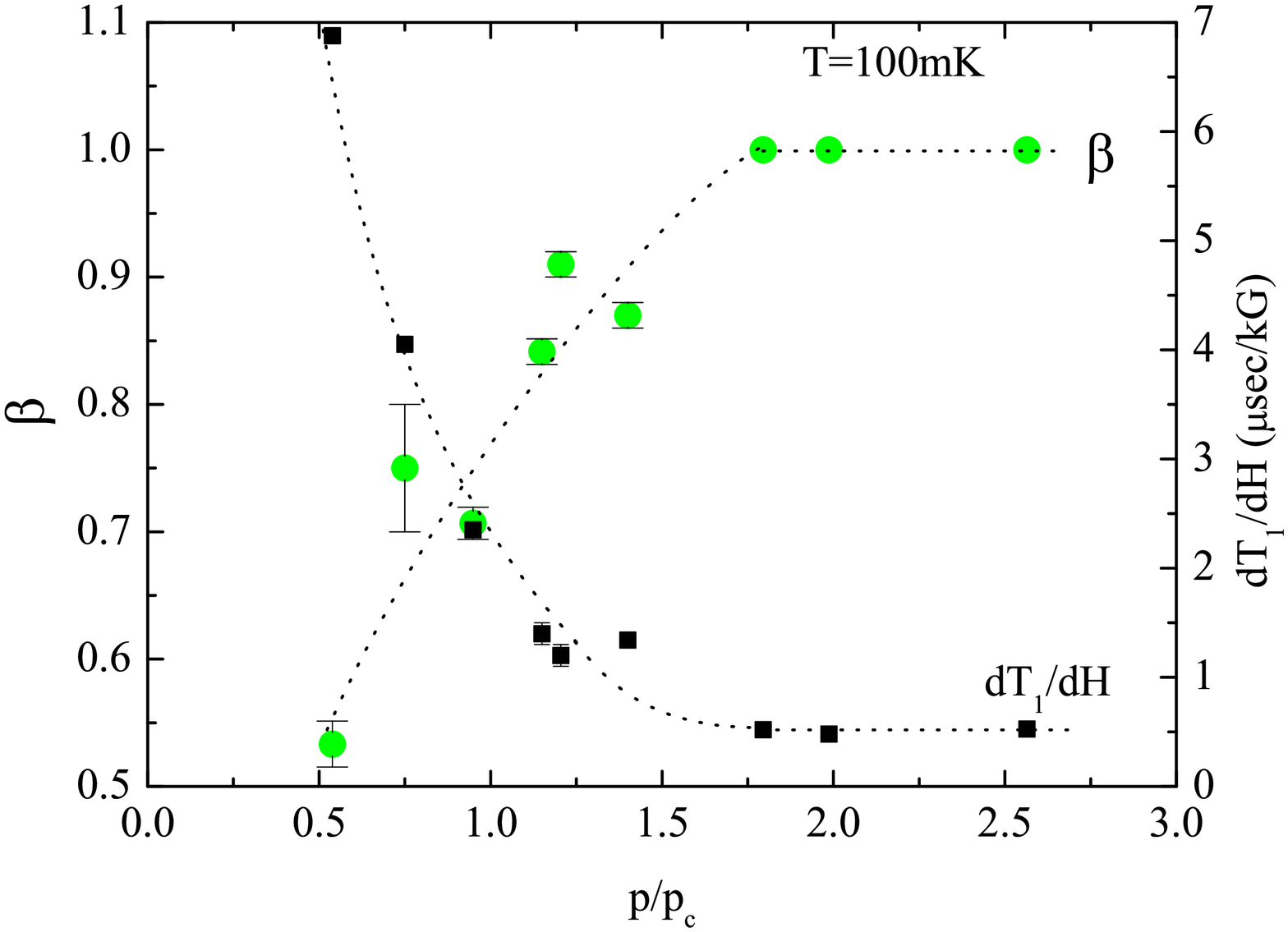}%
\caption{$~dT_{1}/dH$, and $\beta$ vs. $p/p_{c}$ where $p_{c}=0.39$. The
dotted lines are guides to the eye.}%
\label{BandT1vsp}%
\end{center}
\end{figure}

Analysis of the slope of $T_{1}(H)$ at low fields ($dT_{1}/dH$), and the
exponent $\beta$ from Eq.~\ref{FitFunc} as a function of $p$ reveals more
clearly the importance of interactions and percolation. In Fig.
\ref{BandT1vsp} we plot these parameters vs. $p/p_{c}$. The dotted lines are
guides to the eye. Well above $p_{c}$ the $\beta=1$, namely, the muon
relaxation is a pure exponential one, but as $p$ decreases the nature of the
muon relaxation changes gradually and becomes non-exponential with $\beta=0.5$
well below $p_{c}$. Similarly $dT_{1}/dH$ changes smoothly across the
percolation threshold. Neither parameters depend critically on $p_{c}$. We
therefore conclude that while the dynamical fluctuation are sensitive to
interactions, they are not governed by percolating spins but rather by smaller
groups of spins.

\begin{figure}
[ptb]
\begin{center}
\includegraphics[
trim=2.074356in 3.689423in 0.613109in 0.392727in,
height=2.9948in,
width=1.9804in
]%
{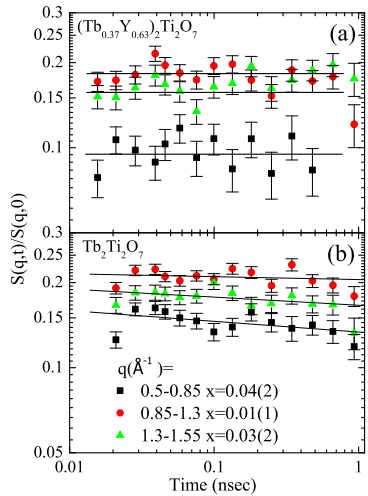}%
\caption{Neutron Spin Echo measurements in two (Tb$_{p}$Y$_{1-p}$)$_{2}%
$Ti$_{2}$O$_{7}$ samples presented on a log-log scale. The solid lines are
linear fits ($x$ is the slope).}%
\label{NSE}%
\end{center}
\end{figure}

Next we present the NSE data on the same polycrystalline pure ($p=1$) and
diluted ($p=0.37$) samples as in the $\mu$SR experiment. These experiments
were performed at the Institut Laue-Langevin, France. The data from the pure
sample only was briefly discussed in Ref.~\cite{JasonNSE} and included here,
with further analysis, for completion. In NSE experiments, which are reviewed
in Ref.~\cite{MezeiNSERev}, a polarized neutron beam is used. The polarization
is perpendicular to the magnetic field $H$, which is directed along the
neutron's momentum. The neutron spins therefore precess, and the difference in
the number of rotations in two identical magnetic fields before and after the
sample is a direct measure of a velocity change (hence, the energy transfer)
that the neutron suffered in the scattering process. By varying the field one
controls the Fourier time $t=1/{}^{n}\gamma H$, and the NSE method measures
the scattering function $S(\mathbf{q},t)$ in the time domain directly, in
addition to the spatial domain ($\mathbf{q}$). The results are presented in
Fig.~\ref{NSE} panels (a) and (b) after summing over several $\mathbf{q}$'s to
reduces statistical error without affecting the $\mathbf{q}$ resolution
significantly. This is possible since the magnetic signal is very broad in
$\mathbf{q}$ \cite{JasonNSE} and NSE is a very poor $\mathbf{q}$ resolution
technique. The relaxation is found to have two components: A fast one, which
cannot be observed by the NSE experiment but leads to an immediate reduction
of the normalized intermediate scattering function $S(\mathbf{q}%
,t)/S(\mathbf{q},0)$, and a slower one. In the pure sample (panel b) the
relaxation of the slow component is consistent with a power law decay since it
is linear on a log-log plot. In the diluted sample (panel a) no relaxation
could be observed. The general evolution of the slow relaxing part of
$S(\mathbf{q},t)$ with doping is consistent with the $\mu$SR experiment. The
more diluted the sample the slower the fluctuations are and the closer this
fraction of the spins to static behavior. While this experiment does not
elucidate on role of $p_{c}$ it serves as a basis for a deeper understanding
of the $\mu$SR data, especially the linear field dependence of the muon
$T_{1}$, which is very unusual.%

\ In all magnets with exponentially decaying correlation function, $T_{1}$ is
a linear function of $H^{2}$ in the low field limit ($g\mu_{B}H\ll J$)
relevant in the present case \cite{KerenPRB01}. Using the NSE results we
speculate on the origin of this linear field dependence and its evolution with
doping. We consider only the slow relaxing part of the NSE data, since only
this part is relevant in the $\mu$SR time window, and we make the assumption
that
\begin{equation}
S(\mathbf{q},t)=F(\mathbf{q})\Phi(t) \label{SqSt}%
\end{equation}
where $F(\mathbf{q})$ is not specified,
\begin{equation}
\Phi(t)=\frac{\tau_{e}^{x}}{(t+\tau_{e})^{x}}, \label{Phioft}%
\end{equation}
$x$ is the power of the correlation function, and $\tau_{e}$ is an early time
cutoff introduced to normalize the correlation at $t=0$. However, $\tau_{e}$
is so small that it can not be observed experimentally. It is therefore
ignored in the denominator. Under this assumption the slope of the lines in
Fig.~\ref{NSE} is $x$. The linear fit in panel (b) gives the $x$ values shown
in the figure. All these values are consistent with $x=0.02$ with no $q$
dependence. In the pure system, where the muon spin relaxation is exponential,
$T_{1}^{-1}$ is determined by the Fourier transform of the auto correlation
function $\left\langle \mathbf{S}_{i}(t)\mathbf{S}_{i}(0)\right\rangle $ at
the frequency $f=\frac{\gamma_{\mu}}{2\pi}H$ \cite{KerenPRB01}. This
correlation function is related to the sum over $\mathbf{q}$ of $S(\mathbf{q}%
,t)$ \cite{KerenHI94}. Therefore, under the assumption of Eq.~\ref{SqSt} we
have
\begin{equation}
\frac{1}{T_{1}(H)}=2\Delta^{2}\int_{0}^{\infty}\Phi(t)\cos(\gamma_{\mu}%
H\tau)d\tau. \label{T1ofH}%
\end{equation}
where $2\Delta^{2}=\gamma_{\mu}^{2}\left\langle \mathbf{B}_{\bot}{}%
^{2}\right\rangle $ is the RMS of the instantaneous transverse (to $H$) field
distribution at the muon site (from the slowly fluctuating part only). This
gives
\begin{equation}
\frac{1}{T_{1}}=\frac{2\Delta^{2}(2\pi f\tau_{e})^{x}}{2\pi f}\Gamma
(1-x)\sin\left(  \frac{\pi x}{2}\right)  . \label{FinalT1ofH}%
\end{equation}
In our experiment $x\rightarrow0$, therefore \cite{Comment}
\begin{equation}
T_{1}(H)\simeq\frac{2f}{x\Delta^{2}}. \label{SmallxLimit}%
\end{equation}
Thus, on the basis of Eqs.~\ref{SqSt} and \ref{Phioft} we can explain the
linear field dependence of the muon $T_{1}$. We associate the fact that
$T_{1}>0$ even when $H=0$ with the presence of static fields from other
sources such nuclear moments for example.%

To increase confidence in Eq.~\ref{SmallxLimit} we check if reasonable values
for the theoretical parameters produce the observed $T_{1}$. For example if
$\tau_{e}\sim10^{-13}$~$\sec$, $\mathbf{B}_{\bot}$ and $H$ on the order of
$100$~G ($2\pi f\sim\Delta\sim10$~MHz) and $x=0.02$, we have $(2\pi f\tau
_{e})^{x}\sim1$ and $T_{1}\sim1$~$\mu\sec$, on the order of the measured
value. We thus believe that Eq.~\ref{SmallxLimit} describes our $\mu$SR
results appropriately.

This allows us to conclude further that steeper $T_{1}(H)$ implies smaller
$x\Delta^{2}$. We argue that it is mostly $x$ that is changing with decreasing
$p$ and not $\Delta$. First, as observed in the NSE experiment, $x$ decreases
with decreasing $p$. Second, the drastic change in $dT_{1}/dH$ near $p_{c}$
cannot be due to a dramatic change in $\Delta$. This parameter, which measures
the local field at the muon site, certainly depends on $p$, but it is not
sensitive to whether there is, or is not, percolation in the system.
Therefore, the slowing down of spin fluctuations with decreasing $p$ is
manifested by a decreasing $x$, namely, the power law decay becomes closer to
a constant.

To summarize, our measurements clearly classify the pyrochlore (Tb$_{p}%
$Y$_{1-p}$)$_{2}$Ti$_{2}$O$_{7}$ as a cooperative paramagnet. The name
\textquotedblleft paramagnet\textquotedblright\ is justified by the plateau in
$1/T_{1}$ at low temperature. The adjective \textquotedblleft
cooperative\textquotedblright\ is appropriate since the spin dynamics depend
on the coupling between magnetic moments. However, whether the spin system
percolate or not is not relevant. For comparison in the kagom\'{e} based
compound SrCr$_{9p}$Ga$_{12-9p}$O$_{19}$ spin dynamic was shown to be
completely impartial to $p_{c}$ \cite{KerenPRL00}. We also show that the
correlation of spins in this magnet has a power law decay with a very small
power $x$. This power gets even smaller when the magnetic coverage decreases
resulting in slowing down the spin fluctuations. Since the Fourier transform
of $t^{-x}$ behaves asymptotically like $f^{x-1}$ the fluctuation spectrum
practically has $1/f$ behavior in the frequency range $0.5$~MHz to $50$~GHz.

We would like to thank the ISIS facilities for their kind hospitality and
continuing support of this project. This work was funded by the Israel - U. S.
Binational Science Foundation, the EU-TMR program, and the NATO Collaborative
Linkage Grant, reference number PST.CLG.978705. JSG's work at Brookhaven is
supported by Division of Material Sciences, U.S. Department of Energy under
contract DE-AC02-98CH10886.

\end{document}